\documentclass[12pt]{iopart}
\usepackage{epsfig}
\usepackage{graphicx}
\usepackage{psfig}

\begin{document}

\title[Mesoscopic thermodynamics of stationary states]{Mesoscopic
thermodynamics of stationary states}
\author{I Santamar\'{\i}a-Holek \dag, J M Rub\'{\i} \ddag, A P\'{e}%
rez-Madrid \ddag}
\address{\dag Facultad de Ciencias, Universidad Nacional Aut\'{o}noma de M\'{e}xico.\\
Circuito exterior de Ciudad Universitaria. 04510, D. F., M\'{e}xico}

\address{\ddag Facultad de F\'{\i}sica, Universitat de Barcelona. \\
Av. Diagonal 647, 08028, Barcelona, Spain} \ead{ivan@graef.fciencias.unam.mx}

\begin{abstract}
A thermodynamics for systems at a stationary state is formulated.
It is based upon the assumption of the existence of local
equilibrium in phase space which enables one to interpret the probability
density and its conjugated nonequilibrium chemical potential as mesoscopic
thermodynamic variables. The probability current is obtained from the
entropy production related to the probability diffusion process and leads to
the formulation of the Fokker-Planck equation. For the case of a gas of
Brownian particles under steady flow in the dilute and concentrated regimes
we derive nonequilibrium equations of state.
\end{abstract}

\maketitle

\section{Introduction}

Systems driven outside equilibrium by the intervention of an external
driving force frequently show  a peculiar behaviour related to their
relevant quantities not observed when they are close to equilibrium. The
dependence of thermodynamic quantities on external gradients \cite{kawasaki}-%
\cite{lutsko-dufty} and the long-range \cite{longrange} and aging \cite%
{debenedetti} behaviours of the correlation functions are examples
illustrating this very common characteristic \cite{oliveira}-\cite{hassager1}.
The explanation of those peculiarities has been the subject of
intense activity in recent years. A common feature observed is that a
description performed by simply proceeding through the extension of
equilibrium concepts to an out-of-equilibrium situation does not always lead
to a correct characterization of the state of the system \cite{evans,matin}.
Such an approach can only be justified when the system is close to
equilibrium in a local equilibrium state in which the relevant variables are
the locally-conserved hydrodynamic fields \cite{degroot}. In any other
circumstance taking place when other variables become relevant, one has to
propose a more general scenario able to provide a complete description of
the system.

Our purpose in this paper is precisely to analyze one case
presenting those typical features: a 'gas' of Brownian particles under a
shear flow \cite{nosotrosPRE}. We will propose a new thermodynamic theory
aimed at the characterization of the system at a stationary state. The
theory is based upon a broader interpretation of the concept of local
equilibrium at the mesoscale \cite{vilar}-\cite{stokes-einstein} which
enables us to analyze irreversible processes taking place in time scales
where fluctuations become manifest. We will derive the
nonequilibrium equation of state for the pressure tensor as a function of
the shear rate in the dilute and concentrated regimes finding the
non-analytic dependences observed in simple and complex fluids \cite%
{kawasaki,ernst,lutsko-dufty,hassager1}.

The paper is organized as follows. In Sec. \textbf{2}, we derive the
Fokker-Planck equation describing the dynamics of a dilute gas of Brownian
particles beyond the point particle approximation and obtain the
hydrodynamic equations from the hierarchy of evolution equations for the
moments of the probability distribution. In section \textbf{3},
expressions of the mobility and transport coefficients depending on
frequency and shear rate are derived from hydrodynamics through the Fax\'{e}%
n theorem. Non-equilibrium equations of state valid in the dilute limit are
proposed in Section \textbf{4} whereas Section \textbf{5} is devoted to the
case of higher concentration. Finally, in the discussion section we
summarize our main results.

\section{Thermodynamics of a 'gas' of Brownian particles in external flow}

We consider a 'gas' of Brownian particles in contact with a heat bath %
and subjected to conditions creating a velocity field $\vec{v}_{0}=\vec{v}%
_{0}(\vec{r},t)$. The dynamic description of the particles is accomplished
by means of the single-particle probability density $P(\vec{r},\vec{%
u},t)$, where $\vec{r}$ is the position of a particle and $\vec{u}$ its
instantaneous velocity. This probability is normalized to
the number of particles and obeys the conservation law
\begin{equation}
\frac{\partial }{\partial t}P+\nabla \cdot \left( \vec{u}P\right) =-\frac{%
\partial }{\partial \vec{u}}\cdot \left( P\vec{v}_{\vec{u}}\right) ,
\label{continuidadP}
\end{equation}%
where $P\vec{v}_{\vec{u}}$\ represents an unknown probability
current in $\vec{u}$-space and $\nabla =\frac{\partial }{\partial \vec{r}}$.

The explicit expression of the current $P\vec{v}_{\vec{u}}$, can be
found by analyzing the dissipation taking place in $\vec{u}$-space,
or equivalently, the entropy production. To this end, we
will propose a mesoscopic thermodynamic description based on the assumption
of local equilibrium in phase space leading us to formulate the
Gibbs equation
\begin{equation}
T\rho \delta s=\rho \delta e-\int \mu \delta Pd\vec{u},  \label{Gibbs ec}
\end{equation}%
in which the entropy depends on the probability distribution function
instead of on the density of mass. Here $\rho (\vec{r},t)$, $s(\vec{r},t)$, $%
e(\vec{r},t)$ and $\mu (\vec{r},t)$ are the density, the entropy, the
internal energy and the nonequilibrium chemical potential per unit of mass
and $T$ the temperature, assumed constant. The density field is
defined as
\begin{equation}
\rho (\vec{r},t)=m\int P(\vec{u},\vec{r},t)d\vec{u},  \label{ro}
\end{equation}%
where $m$ is the mass of the particle, and the velocity field $\vec{%
v}(\vec{r},t)$ is given through the first moment 
\begin{equation}
\rho \vec{v}(\vec{r},t)\equiv m\int \vec{u}P(\vec{u},\vec{r},t)d\vec{u}.
\label{def.momentum}
\end{equation}

The Gibbs equation formulated in (\ref{Gibbs ec}) must be
compatible with the Gibbs entropy \cite{degroot} 
\begin{equation}
s(t)=-k_{B}\int c_{\vec{u}}\ln \frac{P}{P_{le}}d\vec{u}+s_{le}.
\label{p. gibbs}
\end{equation}%
Here $k_{B}$ is the Boltzmann constant, $c_{\vec{u}}=\frac{P}{\rho }$ the
mass fraction of the Brownian particles, $s_{le}$ the local
equilibrium entropy and $P_{le}$ the probability density of the %
local equilibrium reference state given by \cite{nosotrosPRE}
\begin{equation}
P_{le}=e^{\frac{m}{k_{B}T}[\mu _{le}-\frac{1}{2}(\vec{u}-\vec{v}_{0})^{2}]},
\label{Plecolloid}
\end{equation}%
where $\mu _{le}$ is the chemical potential at local equilibrium, and $\frac{%
1}{2}(\vec{u}-\vec{v}_{0})^{2}$ the kinetic energy per unit of mass %
corresponding to the relative motion of the particles with respect to the
flow. Making variations with respect to $P$ in Eq. (\ref{p. gibbs})
and comparing the result with (\ref{Gibbs ec}) we infer the value of the
chemical potential \cite{nosotrosPRE} 
\begin{equation}
\mu =\frac{k_{B}T}{m}\ln P+\frac{1}{2}(\vec{u}-\vec{v}_{0})^{2}.  \label{mu}
\end{equation}

The entropy production follows by taking the time derivative of Eq.
(\ref{p. gibbs}) and by using Eqs. (\ref{continuidadP}), (\ref%
{Gibbs ec}), (\ref{Plecolloid}), and the corresponding equation for $c_{\vec{%
u}}$ \cite{nosotrosPRE}. After integrating by parts and
assuming that the probability current vanishes at the boundaries, one
obtains 
\begin{equation}
\sigma =-\frac{m}{T}\int P\vec{v}_{\vec{u}}\cdot \frac{\partial \mu }{%
\partial \vec{u}}d\vec{u}-\frac{m}{T}\int \vec{J}\cdot \nabla \left[ \frac{1%
}{2}(\vec{u}-\vec{v}_{0})^{2}\right] -\frac{m}{T}\int \vec{J}_{0}\cdot \vec{F%
}d\vec{u},  \label{sigma1part}
\end{equation}%
where $\vec{F}=\frac{d\vec{v}_{0}}{dt}$ is the force per unit of
mass exerted by the flow on the particle. In deriving Eq. (\ref%
{sigma1part}), we have also used the balance equation for the
internal energy $e$, assumed isothermal conditions, neglected the energy
dissipation arising from viscous heating \cite{degroot} and defined $\frac{d%
}{dt}\equiv \frac{\partial }{\partial t}+\vec{v}\cdot \nabla $. The first
contribution to the entropy production (\ref{sigma1part}) is related
to the diffusion process in $\vec{u}$-space. The second %
originates from the relative current $\vec{J}=(\vec{u}-\vec{v})P$, %
whereas the third, involving the current $\vec{J}_{0}=(\vec{u}-\vec{v}_{0})P
$ comes from the dependence on time of the external velocity field $%
\vec{v}_{0}(\vec{r},t)$ \cite{nosotrosPRE}.

According to the Onsager theory \cite{degroot}, we may now
establish couplings between fluxes and forces of the same tensorial order
appearing in Eq. (\ref{sigma1part}). In particular, the diffusion
current in $\vec{u}$ is given by
\begin{equation}
P\vec{v}_{\vec{u}}=-\vec{\vec{\xi}}P\cdot \frac{\partial \mu }{\partial \vec{%
u}}-\vec{\vec{\epsilon}}P\cdot (\vec{u}-\vec{v}_{0})\cdot \nabla \vec{v}_{0}+%
\vec{\vec{\zeta}}\cdot \vec{F}P,  \label{Pv}
\end{equation}%
where we have introduced the coefficients $\vec{\vec{\xi}}$, $\vec{%
\vec{\epsilon}}$ and $\vec{\vec{\zeta}}$ which may, in general, be
functions of position and time \cite{nonmarkov}. Their dependence
on position accounts for spatial inhomogeneities whereas the
dependence on time introduces memory effects \cite{oliveira,nonmarkov,adelman}.

By substituting now Eq. (\ref{Pv}) into (\ref{continuidadP}), one
arrives at the Fokker-Planck equation 
\begin{equation}  \label{GFPE}
\frac{\partial }{\partial t}P+\nabla \cdot \left( \vec{u}P\right) =\frac{%
\partial }{\partial \vec{u}}\cdot \left[ \vec{\vec{\beta }} \cdot (\vec{u}-\vec{v}_{0}) P+\frac{k_{B}T}{m}\vec{\vec{\xi }}\frac{\partial P}{\partial
\vec{u}}\right] -\frac{\partial }{\partial \vec{u}}\cdot \left[ \vec{\vec{%
\zeta }}\cdot \vec{F}P\right] ,
\end{equation}
where we have introduced the friction tensor $\vec{\vec{\beta }} $ related
to $\vec{\vec{\xi }} $ through $\vec{\vec{\beta }}=\vec{\vec{\xi }}+\vec{%
\vec{\epsilon }}\cdot \nabla \vec{v}_{0}$.

In this equation which describes the dynamics of the system
beyond the point particle approximation, the inertial and surface
forces the bath exerts on the particle during its motion are associated to
the terms $\vec{\vec{\zeta}}\cdot \vec{F}$ and $\vec{\vec{\epsilon}}\cdot
\nabla \vec{v}_{0}$, respectively.

From the Fokker-Planck equation, we can directly derive the
evolution equations for the moments of the distribution
which correspond to the hydrodynamical fields \cite{nosotrosPRE}.
In particular, the mass conservation equation 
\begin{equation}
\frac{\partial \rho }{\partial t}=-\nabla \cdot \rho \vec{v},
\label{contmasa}
\end{equation}%
can be obtained by taking the time derivative of equation (\ref{ro}), using (%
\ref{GFPE}) and integrating by parts. By following a similar procedure, from
Eq. (\ref{def.momentum}) we may derive the momentum conservation law
\begin{equation}
\rho \frac{d\vec{v}}{dt}+\nabla \cdot \!\!\,\,\vec{\vec{\mathrm{P}}}%
^{k}\!\!\,\,=-\rho \vec{\vec{\beta}} \cdot (\vec{v}-\vec{v}_{0})+\rho \vec{%
\vec{\zeta}}\cdot \vec{F},  \label{momentum}
\end{equation}%
in which we have introduced the kinetic part of the pressure tensor $\,\vec{%
\vec{\mathrm{P}}}^{k}\!\!(\vec{r},t)$ 
\begin{equation}
\!\!\,\,\vec{\vec{\mathrm{P}}}^{k}\!\!\,\,(\vec{r},t)=m\int (\vec{u}-\vec{v}%
)(\vec{u}-\vec{v})P(\vec{u},\vec{r},t)d\vec{u},  \label{Puu}
\end{equation}%
whose evolution equation is
\begin{equation}
\frac{d}{dt}\!\!\,\,\vec{\vec{\mathrm{P}}}^{k}\!\!\,\,+2\left( \!\!\,\,\vec{%
\vec{\mathrm{P}}}^{k}\!\!\,\,\cdot \tau ^{-1}\right) ^{s}=\frac{2k_{B}T}{m}%
\rho \vec{\vec{\xi}}^{s}.  \label{balanceP}
\end{equation}%
Here, the exponent $s$ refers to the symmetric part and the matrix
of relaxation times $\vec{\vec{\tau}}$ is defined by 
\begin{equation}
\vec{\vec{\tau}}=\left[ \vec{\vec{\beta}}+\nabla \vec{v}+\frac{1}{2}(\nabla
\cdot \vec{v})\vec{\vec{1}}\right] ^{-1}.  \label{tao1}
\end{equation}%
In Eq. (\ref{balanceP}), we have neglected the contribution of the third
centered moment and higher order moments related to the
Burnett and super-Burnett approximations since their characteristic
relaxation times are much smaller than those appearing in Eqs. (\ref%
{contmasa})-(\ref{balanceP}) \cite{nosotrosPRE}. It is important to stress
that Eq. (\ref{balanceP}) constitutes a general expression from which one
may obtain the constitutive equation for $\!\!\,\,\vec{\vec{\mathrm{P}}}%
^{k}\!\!\,\,$ which may incorporate memory effects.

At times $t\gg \beta _{ij}^{-1}$, the particle enters the diffusion
regime governed by the corresponding Smoluchowski equation for $\rho (\vec{r%
},t)$ \cite{nosotrosPRE}. In this regime, Eq. (\ref{balanceP}) yields
\begin{equation}
\!\!\,\,\vec{\vec{\mathrm{P}}}^{k}\!\!\,\,\simeq \frac{k_{B}T}{m}\rho \vec{%
\vec{1}}-\left[ \left( \vec{\vec{\eta}}_{B}+\vec{\vec{\eta}}_{H}\right)
\cdot \nabla \vec{v}_{0}\right] ^{s},  \label{Pvv approx}
\end{equation}%
where we have assumed that for sufficiently long times, $\vec{v}\approx \vec{%
v}_{0}+O(\nabla \ln \rho )$ and have been defined the Brownian $\vec{\vec{%
\eta}}_{B}$ and hydrodynamic $\vec{\vec{\eta}}_{H}$ viscosities 
\begin{equation}
\vec{\vec{\eta}}_{B}=\frac{k_{B}T}{m}\rho \vec{\vec{\beta}}%
^{-1}\,\,\,\,\,\,\,and\,\,\,\,\,\,\,\,\,\vec{\vec{\eta}}_{H}=\frac{k_{B}T}{m}%
\rho \vec{\vec{\beta}}^{-1}\cdot \vec{\vec{\epsilon}}.  \label{eta_{B}}
\end{equation}%
The former originates from the stresses caused by the Brownian
particle in its motion whereas the latter is a consequence of the finite
size of the particle which introduces inertial effects. Eq. (\ref{Pvv
approx}) will constitute the starting point for the derivation of the
 nonequilibrium equations of state of the gas.

\section{The generalized Fax\'{e}n theorem}

The explicit form of the coefficients $\vec{\vec{\beta}}$, $\vec{\vec{%
\epsilon}}$ and $\vec{\vec{\zeta}}$, can be obtained by making use of the Fax%
\'{e}n theorem \cite{mazur-bedo} giving the force exerted by the
fluid on a particle of finite size, which has been generalized to \
the case of a shear flow \cite{agus-miguelshear}. \ For simplicty,
we will neglect the contribution due to Brownian motion in Eqs. (\ref%
{momentum}) and (\ref{Pvv approx}), and assume that $\vec{\vec{\beta}}%
=\beta \vec{\vec{1}}$, $\vec{\vec{\epsilon}}=\epsilon \vec{\vec{1}}$ and $%
\vec{\vec{\zeta}}=\zeta \vec{\vec{1}}$.

By substituting Eq. (\ref{Pvv approx}) into (\ref{momentum}) after
neglecting Brownian contributions, simplifying for $\rho $ and taking the
Fourier transform of the resulting equation, one obtains 
\begin{equation}
\vec{v}(\omega )=G(\omega )\left[ \beta \vec{v}_{0}(\omega )+\frac{k_{B}T}{m}%
\rho \beta ^{-1}\epsilon \nabla ^{2}\vec{v}_{0}(\omega )-i\omega \zeta \vec{v%
}_{0}(\omega )\right] ,  \label{balancev general fourier}
\end{equation}%
where we have defined the propagator of the velocity $G\equiv
\left( -i\omega +\beta \right) ^{-1}$, and $\vec{v}(\omega )\equiv (2\pi )^{-%
\frac{3}{2}}\int \vec{v}e^{-\imath \omega t}dt$. The quantities $%
\vec{v}_{0}(\omega )$ and $\nabla ^{2}\vec{v}_{0}(\omega )$ appearing in Eq.
(\ref{balancev general fourier}) can be expressed as a linear combination
of the averages of the unperturbed velocity over the surface and the volume
of the particle defined respectively as
\begin{equation}
\overline{\vec{v}_{0}}^{S}(\omega )=\frac{1}{4\pi a^{2}}\int \vec{v}%
_{0}(\omega )dS\cong \vec{v}_{0}(\omega )+\frac{a^{2}}{3!}\nabla ^{2}\vec{v}%
_{0}(\omega ),  \label{promedio superficie}
\end{equation}%
and 
\begin{equation}
\overline{\vec{v}_{0}}^{V}(\omega )=\frac{3}{4\pi a^{3}}\int \vec{v}%
_{0}(\omega )d\vec{r}\cong \vec{v}_{0}(\omega )+\frac{a^{2}}{10}\nabla ^{2}%
\vec{v}_{0}(\omega ),  \label{promedio volumen}
\end{equation}%
where $a$ is the radius of the particle and the derivatives of the
velocity are evaluated at the center of mass. Using these relations, Eq. (%
\ref{balancev general fourier}) can be written as 
\begin{eqnarray}
\vec{v}(\omega )=G(\omega )\left[ \frac{k_{B}T}{m}\frac{15}{a^{2}}\beta
^{-1}\left( 1-\epsilon \right) -\frac{3}{2}\beta +\frac{3}{2}\zeta i\omega %
\right] \overline{\vec{v}_{0}}^{S}(\omega )   \nonumber \\
\,\,\,\,\,\,\,\,\,\,\,\,\,\,\,\, -G(\omega )\left[ \frac{k_{B}T}{m}\frac{15}{a^{2}}\beta ^{-1}\left(
1-\epsilon \right) -\frac{5}{2}\beta +\frac{5}{2}\zeta i\omega \right] \cdot 
\overline{\vec{v}_{0}}^{V}(\omega ),  \label{Faxen generalizado}
\end{eqnarray}%
which coincides with Fax\'{e}n's theorem obtained from hydrodynamics
\cite{mazur-bedo} if we make the following identifications
\begin{equation}
\beta =\beta _{0}\left( 1+a\alpha _{\omega }+\frac{1}{9}a^{2}\alpha _{\omega
}^{2}\right) ,  \label{beta fax}
\end{equation}%
\begin{equation}
\epsilon =\frac{1}{6}\frac{m}{k_{B}T}a^{2}\beta _{0}^{2}\left( 1+2a\alpha
_{\omega }+\frac{59}{45}a^{2}\alpha _{\omega }^{2}\right) ,
\label{Epsilon-fax}
\end{equation}%
and $\zeta =\frac{\rho _{p}}{\rho _{f}}$, where $\beta _{0}$ is the Stokes
friction coefficient, $\alpha _{\omega }\equiv \sqrt{\frac{-i\omega }{\nu }}$
is the inverse of the penetration length of the perturbation, $\nu $ the kinematic
viscosity of the solvent and $\rho _{p}$ and $\rho_{f}$ are the particle and fluid densities.

The case of a stationary flow $\vec{v}_{0}=\vec{v}_{0}(\vec{r})$
and zero-frequency can also be analyzed by simply considering the
average over the surface of the particle since in this case inertial effects
vanish. The mobility tensor $\vec{\vec{\beta}}^{-1}$ can also in
this case be compared with the one calculated from hydrodynamics
\cite{agus-miguelshear} 
\begin{equation}
\vec{\vec{\beta}}^{-1}=\beta _{0}^{-1}(1-\vec{\vec{M}}a\alpha _{\gamma }),
\label{beta-de-gama}
\end{equation}%
One also obtains 
\begin{equation}
\vec{\vec{\epsilon}}=\frac{1}{6}\frac{m}{k_{B}T}a^{2}\left\{ \vec{\vec{\beta}%
}\cdot \vec{\vec{\beta}}-\frac{3}{5}\zeta \vec{\vec{\beta}}\cdot \nabla \vec{%
v}_{0}\right\} ,  \label{epsilonshear}
\end{equation}%
where $\alpha _{\gamma }\equiv \sqrt{\frac{\gamma }{\nu }}$ is the
inverse of the penetration length of the perturbation related to the shear
flow, with $\gamma $ the shear rate and $\vec{\vec{M}}$ a matrix
whose elements depend on the boundary conditions on the surface of the
particle \cite{kallus}.

\section{Nonequilibrium equations of state}

Once the expresions of the transport coefficients have been found,
we will proceed to derive the equations of state for the different elements
of the pressure tensor as functions of the shear rate $\gamma $ in the
zero-frequency case.

Since the gas of Brownian particles constitutes a fluid with a
viscosity coefficient proportional to the diffusiveness $D_{0}=\frac{k_{B}T%
}{6\pi a\eta _{f}}$, with $\eta _{f}$ being the viscosity of the heat bath,
in this case we will introduce  the inverse penetration length $\lambda
_{\gamma }\equiv \sqrt{\frac{\gamma }{D_{0}}}$ \ where we have
replaced the kinematic viscosity $\nu $ of the heat bath by the diffusivity $%
D_{0}$ given by the Stokes-Einstein relation.

In the case of a simple shear along the $x $-axis, the expressions for the
elements of the pressure tensor $\! \! \, \, \vec{\vec{\mathrm{P}}}^{k}\! \!
\, \, $ can be obtained by substituting Eqs. (\ref{beta-de-gama})
and (\ref{epsilonshear}) into the long time limit of Eq. (\ref{balanceP}).
For the diagonal components we obtain 
\begin{equation}  \label{Pxx}
\! \! \, \, \mathrm{P}_{xx}^{k}\! \! \, \, =\frac{k_{B}T}{m}\rho \left[
1-\beta _{xx}^{-1}\epsilon _{xy} \gamma\right] -\beta _{xx}^{-1}\left[ \beta
_{yx}+\gamma \right] \! \! \, \, \mathrm{P}_{xy}^{k}\! \! \, \,
\end{equation}
\begin{equation}  \label{Pyy}
\! \! \, \, \mathrm{P}_{yy}^{k}\! \! \, \, =\frac{k_{B}T}{m}\rho -\beta
_{xx}^{-1}\beta _{xy}\! \! \, \, \mathrm{P}_{xy}^{k}\! \! \, \, ,
\end{equation}
and 
\begin{equation}  \label{Pzz}
\! \! \, \, \mathrm{P}_{zz}^{k}\! \! \, \, =\frac{k_{B}T}{m}\rho .
\end{equation}
The only non-diagonal component different from zero is
\begin{equation}  \label{Pxy}
\! \! \, \, \mathrm{P}_{xy}^{k}\! \! \, \, =-\frac{k_{B}T}{2m}\rho
G^{-1}_{0}\left\{ 1+\epsilon _{yy}-\beta _{xx}^{-1}\beta _{xy}\epsilon
_{xy}\right\} \gamma,
\end{equation}
where we have used Eqs. (\ref{Pxx}) and (\ref{Pyy}) and defined $%
G_{0}\equiv \beta _{xx}-\beta _{xx}^{-1}\beta _{xy}\beta _{yx}-\beta
_{xx}^{-1}\beta _{xy}\gamma $.

By substituting now Eqs. (\ref{beta-de-gama}) and (\ref{epsilonshear}%
) into Eqs. (\ref{Pxx})-(\ref{Pxy}), we obtain the propagator $%
1-M_{xx}a\lambda _{\gamma }$ for the elements of the pressure tensor and
from here, we may conclude that non-Newtonian effects appear at lower shear
rates when increasing the radius of the particle.

We will now proceed to analyze Eqs. (\ref{Pxx})-(\ref{Pxy})
by assuming $\gamma <\frac{D_{0}}{a^{2}}$. In this limit, Eq. (\ref%
{Pxy}) can be approximated up to the second order in $\gamma $ to obtain

\begin{equation}
\!\!\,\,\mathrm{P}_{xy}^{k}\!\!\,\,\simeq -\left[ \eta _{B}+\eta _{H}\right]
\gamma ,  \label{Pxy small}
\end{equation}%
where
\begin{equation}
\eta _{B}=\frac{k_{B}T}{m\beta _{0}}\rho \left[ 1-M_{xx}a\lambda _{\gamma
}+2M_{xy}M_{yx}a^{2}\lambda _{\gamma }^{2}\right] ,  \label{etaBexacta}
\end{equation}%
and 
\begin{equation}
\eta _{H}=\frac{1}{6}\beta _{0}a^{2}\rho \left[ 1+M_{xx}a\lambda _{\gamma
}+\left( M_{xx}^{2}-M_{xy}M_{yx}\right) a^{2}\lambda _{\gamma }^{2}\right] .
\label{etaHexacta}
\end{equation}%
\begin{figure}[tbp]
{}
\par
\centering \mbox{\resizebox*{8cm}{!}{\includegraphics{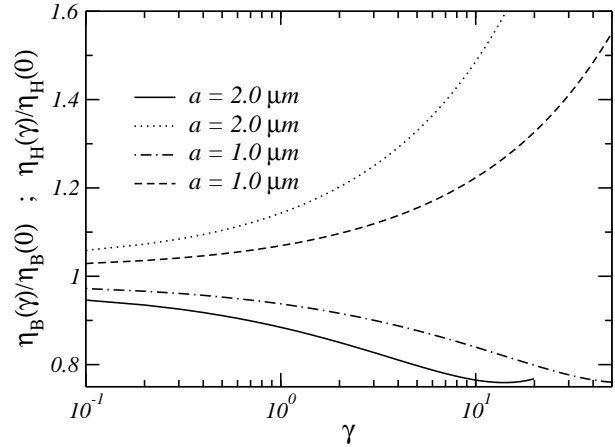}} }
\par
{\footnotesize {\ } \vspace{.8cm} }
\caption{ The Brownian (solid and dashed-dotted lines) and hydrodynamic
(dashed and dotted lines) viscosities as a functions of the shear rate for
different values of the radius of the Brownian particle. This figure shows
the shear thinning and thickening effects. The values of the parameters we
have used are $\protect\beta_{0} \simeq 10^{6}s^{-1}$, $D_{0}\simeq
10^{-6}cm^{2}s^{-1}$ and $\protect\rho \simeq 0.2cm^{3}$. }
\label{ViscByHjuntasvrsgamma}
\end{figure}
In Figure \textbf{1} we present both viscosities normalized to their
value at zero $\gamma $, as a function of the shear rate for different
values of $a$. At low shear rates, Brownian visicosity (dashed-dotted and
solid lines) exhibits shear thinning whereas shear thickening in the
hydrodynamic viscosity (dotted and dahsed lines) arises at higher shear
rates \cite{brady}.

The rheological equations of state follow by substituting Eq. (\ref%
{Pxy small}) with (\ref{etaBexacta}) and (\ref{etaHexacta}) into
Eqs. (\ref{Pxx}) and (\ref{Pyy}) obtaining 
\begin{equation}  \label{Pxx-small}
\! \! \, \, \mathrm{P}_{xx}^{k}\! \! \, \, \simeq \frac{k_{B}T}{m}\rho + %
\left[\frac{k_{B}T}{m\beta_{0}} - \frac{1}{3}\left(\frac{M_{yx}}{M_{xy}}- 
\frac{1}{2}\right)\beta_{0}a^2 \right] M_{xy} a\nu^{-\frac{1}{2}} \rho
\gamma^{\frac{3}{2}},
\end{equation}
and 
\begin{equation}  \label{Pyy-small}
\! \! \, \, \mathrm{P}_{yy}^{k}\! \! \, \, \simeq \frac{k_{B}T}{m}\rho +
\left(\frac{k_{B}T}{m\beta_{0}} + \frac{1}{6} \beta_{0}a^2 \right)M_{xy}
a\nu^{-\frac{1}{2}} \rho \gamma^{\frac{3}{2}},
\end{equation}
from which one concludes that the lowest order contribution
to the normal stress difference is proportional to $\gamma ^{\frac{3}{2}} $. Our
results containing corrections to the viscosity have the same
shear rate dependence as those obtained in Ref. [1] for simple fluids.

\section{Nonequilibrium equations of state for a non-ideal gas}

The thermodynamics of the gas for the case of interacting Brownian
particles can also be formulated as in the dilute case.
Within a mean field approximation in which one assumes that the dynamics is
that of a Brownian particle moving through an effective medium consisting of
the fluid and the remaining particles, the Fokker-Planck equation is \cite%
{stokes-einstein} 
\begin{eqnarray}
\frac{\partial }{\partial t}P+\nabla \cdot \left( \vec{u}P\right) -\nabla %
\left[ \frac{k_{B}T}{m}\ln f\right] \cdot \frac{\partial P}{\partial \vec{u}}
=  \nonumber  \label{GFPE-int} \\
\,\,\,\,\,\,\,\,\,\,\,\,\,\,\,\,\frac{\partial }{\partial \vec{u}}\cdot %
\left[ \vec{\vec{\beta}}\cdot (\vec{u}-\vec{v}_{0})P+\frac{k_{B}T}{m}\vec{%
\vec{\xi}}\cdot \frac{\partial P}{\partial \vec{u}}\right] -\frac{\partial }{%
\partial \vec{u}}\cdot \left[ \vec{\vec{\zeta}}\cdot \vec{F}P\right] .
\end{eqnarray}%
Interactions have been taken into account through the activity coefficient $%
f=\exp [\frac{m}{k_{B}T}\rho ^{-1}p^{int}]$ measuring the deviation of the
Brownian gas with respect to the ideal case, with $p^{int}$ being the excess
of osmotic pressure \cite{stokes-einstein}.

Proceeding along the lines indicated in Sec. \textbf{3}, one may
use Eq. (\ref{GFPE-int}) in order to derive the evolution equation of the
momentum 
\begin{equation}
\rho \frac{d\vec{v}}{dt}=-\rho (\vec{v}-\vec{v}_{0})\cdot \vec{\vec{\beta}}%
-\nabla \cdot \!\!\,\,\vec{\vec{\mathrm{P}}}\!\!\,\,+\left[ \frac{k_{B}T}{m}%
\ln f\right] \nabla \rho +\rho \zeta \cdot \vec{F},  \label{momentum-int}
\end{equation}%
where we have defined the total pressure tensor by $\!\!\,\,\vec{\vec{%
\mathrm{P}}}\!\!\,\,=\!\!\,\,\vec{\vec{\mathrm{P}}}\!\!\,\,^{k}+p^{int}\vec{%
\vec{1}}$. The expression of the kinetic part of the pressure tensor
is given by $\!\!\,\,\mathrm{P}_{xy}\!\!\,\,^{k}\simeq \frac{k_{B}T}{m}\rho -%
\left[ \eta _{B}(\omega )+\eta _{H}(\omega )\right] \gamma $, with
dependences of the viscosities different  from those in Eq (\ref{Pvv approx}%
). In deriving the corresponding expression of $\!\!\,\,\vec{\vec{\mathrm{P}}%
}\!\!\,\,$, we have assumed that the relaxation of the kinetic part of the
pressure tensor is much faster than that of normal
(structural) stresses.

\subsection{Viscoelastic effects}

As in the previous section, the transport coefficients can be
identified by using the Fax\'{e}n theorem. At higher volume fractions $\phi $
of the Brownian gas, one may assume that these coefficients can be expressed
in terms of modified penetration lengths.

In the creeping flow case, the mobility coefficient can be expanded up to
the first order in terms of the modified inverse penetration length 
\begin{equation}
\alpha _{\omega }\equiv \frac{1}{\Delta _{\omega }a}\left( \tau \omega
\right) ^{\delta _{\omega }},  \label{alpha-omega-int}
\end{equation}%
where the scaling factor $\Delta _{\omega }$ and the
exponent $\delta _{\omega }$ can, in general, be functions of the volume
fraction $\phi $, \cite{stokes-einstein}. In Eq. (\ref{alpha-omega-int}), we
have introduced the characteristic diffusion time $\tau \equiv \frac{a^{2}}{%
6D_{0}}$. By using Eq. (\ref{alpha-omega-int}), the corresponding relation
for $\beta _{\omega }$ and $\epsilon _{\omega }$ can be expressed by \cite%
{stokes-einstein} 
\begin{equation}
\beta _{\omega }=\beta _{0}\left( 1+a\alpha _{\omega }\right)
\,\,\,\,\,\,and\,\,\,\,\,\,\epsilon _{\omega }=\frac{1}{6}\frac{m}{k_{B}T}%
a^{2}\beta _{0}^{2}\left( 1+2a\alpha _{\omega }\right) .
\label{beta-epsilon-fax-int}
\end{equation}%
Consequently, the frequency-dependent Brownian viscosity
coefficient is now given by
\begin{equation}
\eta _{B}(\omega )=\frac{k_{B}T}{m\beta _{0}}\rho \left[ 1-\Delta _{\omega
}^{-1}\left( \tau \omega \right) ^{\delta _{\omega }}+\Delta _{\omega
}^{-2}\left( \tau \omega \right) ^{2\delta _{\omega }}\right] ,
\label{etaB-int}
\end{equation}%
and the hydrodynamic viscosity coefficient by 
\begin{equation}
\eta _{H}(\omega )=\frac{1}{12}\beta _{0}a^{2}\rho \left[ 1+\Delta _{\omega
}^{-1}\left( \tau \omega \right) ^{\delta _{\omega }}-\Delta _{\omega
}^{-2}\left( \tau \omega \right) ^{2\delta _{\omega }}\right] .
\label{etaH-int}
\end{equation}%
The first normal stress difference is accordingly
\begin{equation}
\!\!\,\,\mathrm{P}_{xx}\!\!\,\,-\!\!\,\,\mathrm{P}_{yy}\!\!\,\,\simeq \beta
^{-1}\left[ \eta _{B}(\omega )+\eta _{H}(\omega )\right] \gamma ^{2},
\label{Pxx-Pyy-int}
\end{equation}%
and then beomes a function of frequency and volume fraction. At
sufficiently low shear rates, the quadratic dependence on the shear rate in
the first normal stress difference has been reported in the literature \cite%
{todd1,sasa1}. Notice that in the ideal case, similar dependence can
be obtained in this regime.

\subsection{Non-Newtonian effects}

In the stationary flow case, we will assume that the mobility can
be expanded in terms of the \textit{modified} inverse penetration length $%
\Lambda _{\gamma }$ 
\begin{equation}
\Lambda _{\gamma }\equiv \frac{1}{\Delta _{\gamma }a}\left( \tau \gamma
\right) ^{\delta _{\gamma }},  \label{alpha-gamma-int}
\end{equation}%
where $\Delta _{\gamma }$ and $\delta _{\gamma }$ are functions of
the volume fraction $\phi $. The mobility can be expressed as 
\begin{equation}
\beta _{\gamma }^{-1}=\beta _{0}^{-1}(\vec{\vec{1}}-\vec{\vec{M}}a\Lambda
_{\gamma }),  \label{beta-gamma-int}
\end{equation}%
and the explicit expression of $\epsilon _{\gamma }$ can be found by using
Eq. (\ref{epsilonshear}). In this case, the contributions to the
viscosity are given by
\begin{equation}
\eta _{B}(\gamma )=\frac{k_{B}T}{m\beta _{0}}\rho \left[ 1-M_{xx}\Delta
_{\gamma }^{-1}\left( \tau \gamma \right) ^{\delta _{\gamma
}}+M_{xy}M_{yx}\Delta _{\gamma }^{-2}\left( \tau \gamma \right) ^{2\delta
_{\gamma }}\right],  \label{etaB-int-gama}
\end{equation}%
and 
\begin{equation}
\eta _{H}(\gamma )=\frac{1}{6}\beta _{0}a^{2}\rho \left[ 1+M_{xx}\Delta
_{\gamma }^{-1}\left( \tau \gamma \right) ^{\delta _{\gamma }}+\left(
M_{xx}^{2}-M_{xy}M_{yx}\right) \Delta _{\gamma }^{-2}\left( \tau \gamma
\right) ^{2\delta _{\gamma }}\right]  \label{etaH-int-gama}
\end{equation}%
\begin{figure}[tbp]
{}
\par
\centering \mbox{\resizebox*{8cm}{!}{%
\includegraphics{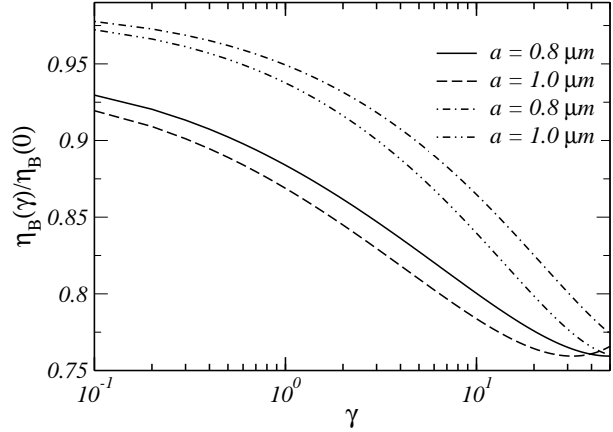}} }
\par
\vspace{.8cm}
\caption{ {\protect\footnotesize {\ Comparison of the normalized Brownian viscosity in
the ideal (dash-dotted lines) and non-ideal cases (dashed and solid lines)
as a function of the shear rate $\protect\gamma $ for two values of $a$. In
the non-ideal case, we have chosen $\Delta _{\protect\gamma }=0.5$ and $%
\protect\delta _{\protect\gamma }=\frac{1}{3}$ which correspond to a density
of particles of $0.46$, \cite{stokes-einstein}. }}}
\label{Viscosidadesrsgamma}
\end{figure}
The shear rate dependent first normal stress difference is now
\begin{equation}
\!\!\,\,\mathrm{P}_{xx}\!\!\,\,-\!\!\,\,\mathrm{P}_{yy}\!\!\,\,\simeq -\frac{%
M_{xy}}{3}\beta _{0}a^{2}\rho \Gamma ^{-1}\tau ^{\delta _{\gamma }}\gamma
^{1+\delta _{\gamma }}.  \label{Pxx-Pyy-int-gamma}
\end{equation}
These results show that the form of the equations of state is
affected by the interactions by modifying the magniutde of the Brownian
viscosity and the dependence of the
exponents on the volume fraction, which is responsible for the loss of universality
in the power law behavior of the viscosity in concentrated systems \cite%
{brady2000}.

Figure \textbf{2} illustrates the differences in shear thinning behaviour
in both the ideal and non-ideal cases. Since in
the general case the inverse penetration length depends on the sum of the
frequency and the shear rate \cite{Zwanzig81}, then we have calculated the value
of the exponent $\delta_{\gamma}$  by using the results of Ref. \cite%
{stokes-einstein} for a density of particles of $\rho \simeq 0.46$. From this figure and Eqs. (\ref{etaB-int-gama}) and (\ref{etaH-int-gama}), one concludes that shear thinning
effects are more pronounced in the non-ideal case at lower values of $\gamma$.

\section{Discussion}

By means of a reformulation of the local equilibrium hypothesis in the
domain of the mesoscale, we have shown that a nonequilibrium thermodynamic
treatment including the presence of fluctuations and of external forces
driving the system to nonequilibrium steady states is possible. Using this
scheme, we have obtained the Fokker-Planck equation which in the case of an
inhomogeneous bath, is coupled to the evolution equations of the fields
characterizing the state of the bath. The evolution equations of the
hydrodynamic fields are derived from the Fokker-Planck equation; their
stationary solutions lead to nonequilibrium equations of state.

We have studied, in particular, the case of a gas of Brownian particles in
contact with a heat bath when the whole system is in motion. The ideal and
non-ideal cases have been analyzed. In both cases, we have obtained
nonequilibrium equations of state showing a dependence of the pressure
tensor of the gas on the frequency and on the shear rate, thus revealing the
presence of viscoelastic and non-Newtonian effects.

In the ideal case, the non-analytical dependence on the shear rate
of the corrections to the first normal stress difference and the viscosities
coincide with those reported in Ref. \cite{kawasaki} for a simple fluid. In
the non-ideal case, nonequilibrium equations of state have been derived by
assuming that the inverse penetration length is similar to that of the ideal
case with an exponent and a scaling factor depending on the volume fraction
of the 'gas' of Brownian particles. Our results show that, at sufficiently low shear
flows, the first normal stress difference is a quadratic function of
the shear rate thus coinciding with the simulations reported in Ref. \cite%
{todd1}. At higher shear flows, non-Newtonian effects are manifested
through a power law dependence on the shear rate of the first normal stress
difference and on the viscosities \cite{hassager1,sasa1}.

Our general framework can be useful in the study of the dynamics of small
systems subjected to driving forces or gradients for which the
presence of fluctuations may play a significant role.

\section{Acknowledgements}

We acknowledge Prof. L. Garc\'{i}a-Col\'{i}n and Dr. D. Reguera for their valuable discussions and comments. I.S.H. acknowledges Drs. R. Rodriguez and C. M\'{a}laga,
and M.C. F. Mandujano for their helpful discussions, and UNAM-DGAPA for economic
support. This work was partially supported by DGICYT of the Spanish Government
under Grant No. PB2002-01267.


\section{References}

\begin{thebibliography}{99}
\bibitem{kawasaki} K. Kawasaki and J. D. Gunton, Phys. Rev. A \textbf{8},
2048 (1973).

\bibitem{ernst} M. H. Ernst, et al., J. Stat. Phys., \textbf{18}, 237 (1978).

\bibitem{lutsko-dufty} J. Lutsko and J. W. Dufty, Phys. Rev. A \textbf{32},
1229 (1985).

\bibitem{longrange} I. Pagonabarraga and J. M. Rub\'{\i}, Phys. Rev. E 
\textbf{49}, 267 (1994).

\bibitem{debenedetti} P. G. Debenedetti and F. H. Stillinger, Nature \textbf{%
410}, 259 (2001).

\bibitem{oliveira} I. V. L. Costa , R. Morgado, M. V. B. T. Lima and
F. A. Oliveira, Europhys. Lett. \textbf{63}, 173 (2003).

\bibitem{hess} O. Hess and S. Hess, Physica A \textbf{207}, 517 (1994).

\bibitem{zatloukal} M. Zatloukal, J. Non-Newtonian Fluid Mech. \textbf{113},
209 (2003).

\bibitem{todd1} J. Ge, G. Marcelli, B. D. Todd and R. J. Sadus, Phys. Rev. E 
\textbf{64}, 021201 (2001).

\bibitem{mellema} B. van der Vorst, et al, Phys. Rev. E \textbf{56}, 3119
(1997).

\bibitem{hassager1} R. B. Bird, C. F. Curtiss, R. C. Armstrong and O.
Hassager, \emph{Dynamics of polymeric fluids. Vols. 1 and 2} (John Wiley \&
Sons, New York, 1987).

\bibitem{evans} H. J. Hanley and D. J. Evans, J. Chem. Phys. \textbf{76},
3225 (1982).

\bibitem{matin} P. J. Daivis and M. L. Matin, J. Chem. Phys. \textbf{118},
11111 (2003).

\bibitem{degroot} S. R. de Groot and P. Mazur, \emph{Non-equilibrium
thermodynamics} (Dover, New York, 1984).

\bibitem{nosotrosPRE} I. Santamar\'{\i}a-Holek, D. Reguera and J. M. Rub%
\'{\i}, Phys. Rev. E \textbf{63} 051106 (2001).

\bibitem{vilar} J. M. Vilar and J. M. Rub\'{\i}, Proc. Natl. Acad. Sci. 
\textbf{98}, 11081 (2001).

\bibitem{gradtemp} A. P\'{e}rez-Madrid, J. M. Rub\'{\i}, P. Mazur, Physica A 
\textbf{212}, 231 (1994).

\bibitem{stokes-einstein} J. M. Rub\'{\i}, I. Santamar\'{\i}a-Holek and A. P%
\'{e}rez-Madrid, J. Phys. Cond. Matter, \textbf{16}, S2047 (2004).

\bibitem{nonmarkov} I. Santamar\'{\i}a-Holek and J. M. Rub\'{\i}, Physica A 
\textbf{326}, 384 (2003).

\bibitem{adelman} S. A. Adelman, J. Chem. Phys. \textbf{64}, 124 (1976).

\bibitem{mazur-bedo} P. Mazur and D. Bedeaux, Physica \textbf{76}, 235
(1974).

\bibitem{agus-miguelshear} A. P\'{e}rez-Madrid, J. M. Rub\'{\i} and D.
Bedeaux, Physica A \textbf{163}, 778(1990).

\bibitem{kallus} S. Kallus, \textit{et al}, Rheol. Acta \textbf{40}, 552
(2001).

\bibitem{brady} G. Bossis and J. F. Brady, J. Chem. Phys. \textbf{91}, 1866
(1989).

\bibitem{sasa1} H. Wada and S. Sasa, Phys. Rev. E \textbf{67}, 065302(R)
(2003).

\bibitem{brady2000} D. R. Foss, and J. F. Brady, J. Rheol. \textbf{44} 629
(2000).

\bibitem{Zwanzig81} R. Zwanzig, Proc. Natl. Acad. Sci. USA, 78, 3296 (1981).
\end{thebibliography}
\end{document}